\documentclass
[prl,twocolumn,superscriptaddress,floatfix,showpacs,10pt]{revtex4}%
\usepackage{amsfonts}
\usepackage{amsmath}
\usepackage{amssymb}
\usepackage{graphicx}%
\begin{document}

\title{Exceptional Form of $D=11$ Supergravity}

\newcommand{\be}{\begin{equation}}
\newcommand{\ee}{\end{equation}}
\newcommand{\ben}{\begin{displaymath}}
\newcommand{\een}{\end{displaymath}}
\newcommand{\bea}{\begin{eqnarray}}
\newcommand{\eea}{\end{eqnarray}}
\newcommand{\nn}{\nonumber}
\newcommand{\non}{\nonumber\\}
\newcommand{\bean}{\begin{eqnarray*}}
\newcommand{\eean}{\end{eqnarray*}}
\newcommand{\beqs}{\begin{eqnarray}}
\newcommand{\eeqs}{\end{eqnarray}}

\author{Olaf Hohm}
\email{olaf.hohm@physik.uni-muenchen.de}
\affiliation{Arnold Sommerfeld Center for Theoretical Physics, Theresienstrasse 37, D-1-80333 Munich, Germany}

\author{Henning Samtleben}
\email{henning.samtleben@ens-lyon.fr}
\affiliation{Universit\'e de Lyon, Laboratoire de Physique, UMR 5672, CNRS, \'Ecole Normale Sup\'erieure de Lyon,
46, all\'ee d'Italie, F-69364 Lyon cedex 07, France}

\begin{titlepage}


\end{titlepage}

\begin{abstract}
Eleven-dimensional supergravity reveals large exceptional symmetries 
upon reduction, in accordance with the U-duality groups 
of M-theory, but their higher-dimensional geometric origin has remained a mystery. 
In this letter, we show that $D=11$ supergravity can be extended to be fully 
covariant under the exceptional groups E$_{n(n)}$, $n=6,7,8$. 
Motivated by a similar formulation of double field theory we introduce an 
extended `exceptional spacetime'. 
We illustrate the construction by 
giving the explicit E$_{6(6)}$ covariant form: the full $D=11$ supergravity, 
in a $5+6$ splitting of coordinates but \textit{without truncation}, 
embeds into an E$_{6(6)}$ covariant  $5+27$ dimensional theory. 
We argue that this covariant form likewise comprises type IIB supergravity.

\end{abstract}

\maketitle

\setcounter{equation}{0}

Little is known about the fundamental formulation of M-theory,
whose low-energy limit is given by 
11-dimensional supergravity~\cite{Cremmer:1978km}.
One illuminating feature is the existence of duality 
symmetries, which relate M-theory to the 10-dimensional superstring theories. 
These symmetries should be as fundamental for the 
formulation of M-theory as diffeomorphism invariance is 
for Einstein's theory of general relativity. 
Intriguingly, the so-called U-duality symmetries comprise 
the exceptional Lie groups  E$_{n(n)}(\mathbb{Z})$~\cite{Hull:1994ys}. 
In the low-energy limit, it has been known for a long time that upon torus 
compactification $D=11$ supergravity gives rise to the continuous 
versions E$_{n(n)}(\mathbb{R})$ \cite{Cremmer:1979up}. Since the early 
1980's this has led to the question: what is it about $D=11$
supergravity that knows about exceptional symmetries? 
It is the purpose of this letter to give  fully E$_{n(n)}$-covariant versions  
of $D=11$ supergravity by employing and generalizing techniques 
from `double field theory' (DFT), an approach that doubles coordinates in order to make the $O(d,d)$ T-duality 
group manifest \cite{Siegel:1993th,Hull:2009mi,Hull:2009zb,Hohm:2010jy,Hohm:2010pp}. 
These formulations show the emergence of 
exceptional symmetries in terms of the higher-dimensional 
geometry and symmetries {\em prior} to any reduction or truncation.

Attempts to understand these `hidden' symmetries in terms of the 
higher-dimensional theory have in fact a long history, at least going back to the 
work of 
de~Wit and Nicolai \cite{deWit:1986mz}, who performed a Kaluza-Klein-like 
decomposition of $D=11$ supergravity in order to exhibit already in eleven dimensions the 
composite local symmetries of the lower-dimensional coset models. 
These formulations did not make the exceptional symmetries manifest, 
and further work in \cite{Koepsell:2000xg} suggested that 
additional coordinates need to be introduced in order to realize 
the exceptional groups.
The idea of such an `exceptional spacetime' has been implemented for a particular 
truncation of $D=11$ supergravity in \cite{Hillmann:2009ci}.
(For more ambitious proposals 
see \cite{West:2001as,Damour:2002cu,West:2003fc}.)
More recently, after the emergence of DFT, a number of papers have succeeded to generalize 
this approach to various U-duality groups, 
see, e.g.,~\cite{Berman:2011jh,Coimbra:2011ky}.  
All these results, however, are restricted to particular truncations of 
$D=11$ supergravity, 
setting to zero the off-diagonal components 
of the metric and of the 3-form, assuming that all fields depend 
only on `internal' coordinates,  and freezing the external metric to be flat Minkowski
up to a possible warp factor.  
This leaves open the question about the 
significance of exceptional symmetries for the full theory. 
The first example of a U-duality covariant formulation of a complete gravity theory
has been obtained in~\cite{Hohm:2013jma} for the `toy-model' of four-dimensional
Einstein gravity. 
By proper Kaluza-Klein type decomposition of fields and extension of the coordinates,
the full theory takes a form that is manifestly covariant under the SL$(2,\mathbb{R})$ Ehlers
symmetry discovered in dimensional reduction more than 50 years ago~\cite{Ehlers:1957}. 
The resulting theory closely resembles DFT when performing the analogous 
Kaluza-Klein type decomposition of fields \cite{Hohm:2013nja}. In the following we apply 
this strategy to $D=11$ supergravity
and embed it into a form that is fully covariant under the exceptional groups E$_{n(n)}$, $n=6,7,8$. 
We argue that this covariant form likewise encodes the type IIB theory~\cite{Schwarz:1983wa}.

For definiteness, we present in detail the case of E$_{6(6)}$ and comment on the other cases below.
Performing a $5+6$ decomposition of the $D=11$ coordinates and embedding the six coordinates
into the fundamental 27-dimensional representation of E$_{6(6)}$, we cast the bosonic sector of 
eleven-dimensional supergravity, without any truncation, into the E$_{6(6)}$-covariant form 
\bea
\label{finalaction}
 S &=&  \int d^5x\, d^{27}Y\,e\, {\cal L}\;, 
\\[0.5ex]
{\cal L}  &\equiv &   \widehat{R}+\frac{1}{24}\,g^{\mu\nu}{\cal D}_{\mu}{\cal M}^{MN}\,{\cal D}_{\nu}{\cal M}_{MN}\nonumber\\[0.5ex]
 &&{}-\frac{1}{4}\,{\cal M}_{MN}{\cal F}^{\mu\nu M}{\cal F}_{\mu\nu}{}^N
 +e^{-1}{\cal L}_{\rm top}-V({\cal M},e) \,.
 \nonumber
\eea
The different terms will be defined in detail below. They resemble the generic structure of non-abelian gauged supergravity
in five dimensions in the embedding tensor formalism~\cite{deWit:2004nw} 
(Einstein-Hilbert, scalar sigma-model, Yang-Mills term, topological term and scalar potential). 
In addition to the 5 coordinates $x^\mu$, 
all fields depend on the 27 extended coordinates $Y^M$, with conjugate derivatives 
$\partial_M$, subject to the E$_{6(6)}$-covariant constraint~\cite{Coimbra:2011ky,Berman:2012vc}
 \be
  d^{MNK}\,\partial_N \partial_K A \ = \ 0\;, \quad  d^{MNK}\,\partial_NA\, \partial_K B \ = \ 0 \,, 
 \label{sectioncondition}
 \ee  
which holds for arbitrary fields and gauge parameters $A,B$. 
This constraint is the `section condition' that is the M-theory 
analogue of the `strong constraint' in DFT, which in turn is a stronger version of the 
level-matching constraint in string theory. 
It can be locally solved by fields depending on only $6$ out of the $Y^M$ coordinates.
With this explicit solution (breaking E$_{6(6)}$ covariance)
the theory (\ref{finalaction}) can be shown to describe the field content  
and dynamics of the full $D=11$ supergravity.
 
Apart from the global E$_{6(6)}$,
the action (\ref{finalaction}) is invariant under generalized ($Y^M$-dependent) diffeomorphisms in the $x^\mu$ coordinates
and in the $Y^M$ coordinates, respectively, of which the latter take the form of non-abelian gauge transformations associated to
the Yang-Mills gauge field. 
In particular, we show that all relative coefficients in (\ref{finalaction}) are determined by 
this generalized gauge- and diffeomorphism invariance. 

We will now introduce the relevant structures in order to define all terms in (\ref{finalaction}). 
The Lie algebra of E$_{6(6)}$ is spanned by generators  $t_{\alpha}$, with the adjoint index $\alpha=1,\ldots, 78$. 
The fundamental  representation ${\bf 27}$ and its dual $\overline{\bf 27}$ will be indicated by indices $M,N=1,\ldots,27$.  
There are two E$_{6(6)}$-invariant tensors, the fully symmetric $d$-symbols $d^{MNK}$ and $d_{MNK}$, 
which we normalize by $d_{MPQ}d^{NPQ} = \delta_M^N$\,.
Generalized Lie derivatives 
with respect to a gauge parameter $\Lambda^M$, acting on a contravariant vector $V^M$ 
of {\em weight} $\lambda$, are defined as \cite{Coimbra:2011ky,Berman:2012vc}
\bea
\begin{split}
\delta_\Lambda V^M  \equiv \;& 
\Lambda^K \partial_K V^M \\
& - 6\, \mathbb{P}^M{}_N{}^K{}_L\,\partial_K \Lambda^L\,V^N
+\lambda\, \partial_P\Lambda^P V^M
\,,
\label{deltaV}
\end{split}
\eea
with the projector onto the adjoint representation
\bea\label{projector}
\mathbb{P}^M{}_N{}^K{}_L
&\equiv& (t_\alpha)_N{}^M (t^\alpha)_L{}^K \\ \nonumber
 &=&
\frac1{18}\,\delta_N^M\delta^K_L + \frac16\,\delta_N^K\delta^M_L
-\frac53\,d_{NLR}\,d^{MKR}
\,.
\eea
These gauge transformations close on fields satisfying (\ref{sectioncondition}), 
$[\delta_{\Lambda_1},\delta_{\Lambda_2}] = \delta_{[\Lambda_2,\Lambda_1]_{\rm E}}$, 
according to the `E-bracket' 
\bea
\big[ \Lambda_2,\Lambda_1 \big]^M_{\rm E}  = 2\Lambda_{[2}^K \partial_K \Lambda_{1]}^M
-10 d^{MNP}d_{KLP}\,\Lambda_{[2}^K \partial_N \Lambda_{1]}^L \;, 
\eea
which is the exceptional E$_{6(6)}$-covariant analogue of the `C-bracket' in DFT. 
The theory (\ref{finalaction}) will be invariant under a local version of  (\ref{deltaV}),
i.e.\ with gauge parameters $\Lambda^M$ that are functions of $Y^M$ and $x^{\mu}$,
and a proper choice of weights for the fields. Accordingly, we
introduce 27 gauge fields $A_{\mu}{}^{M}$ and  define covariant derivatives for the `external' 
5-dimensional derivatives $\partial_{\mu}$
as
\bea
   {\cal D}_{\mu}V^M& \!=\!&  \, \partial_{\mu}V^M-A_{\mu}{}^K\partial_KV^M 
  +\frac{1-3\lambda}{3} \,
  \partial_K A_{\mu}{}^K V^M
\nonumber  \\ 
  &&{} \hspace{-2.9ex}  +V^K\partial_K A_{\mu}{}^M -10d^{MNP}d_{PKL}\partial_N 
  A_{\mu}{}^K V^L \,,
\label{DV}
 \eea
 for a vector of weight $\lambda$\,.
 The  gauge variation of the vector field $A_{\mu}{}^{M}$ is determined by requiring that these covariant derivatives
 transforms covariantly, which in turn implies that 
\be
\delta A_\mu{}^M \equiv {\cal D}_{\mu}\Lambda^M \;, 
\ee
with the weight of $\Lambda^M$ fixed to be $\lambda=\frac13$\,.
The corresponding covariant field strength for $A_{\mu}{}^{M}$ reads 
\bea
{\cal F}_{\mu\nu}{}^M \equiv
 2\partial_{[\mu} A_{\nu]}{}^M 
-\big[A_{\mu}, A_{\nu}\big]^M_{\rm E}
+ 10\,  d^{MNK} \partial_K B_{\mu\nu N}, 
\label{defF}
\eea
where the particular coupling to 27 antisymmetric 2-forms $B_{\mu\nu M}$ is required in order to
achieve gauge covariance, in precise analogy to the tensor hierarchy of 
gauged supergravity~\cite{deWit:2004nw}. 
These 2-forms have their own gauge symmetry with 1-form parameters $\Xi_{\mu M}$
of weight $\lambda=\frac23$\,. 
Together, it may be shown that the field strength (\ref{defF}) transforms covariantly 
according to (\ref{deltaV}) under 
the combined vector and tensor gauge transformations 
 \bea
   \delta A_\mu{}^M &=&  {\cal D}_\mu \Lambda^M -10\,d^{MNK} \partial_K\Xi_{\mu N}\,, \nonumber\\[0.5ex]
   \delta B_{\mu\nu M} &=&  2{\cal D}_{[\mu}\Xi_{\nu]M} 
   +d_{MKL}\Lambda^K{\cal F}_{\mu\nu}{}^{L} \nonumber\\
   &&{}-d_{MKL}\,A_{[\mu}{}^K\, \delta A_{\nu]}{}^L+ {\cal O}_{\mu\nu M}\;,
   \label{deltaAB}
\eea
which are only determined up to terms ${\cal O}_{\mu\nu M}$ satisfying 
 \be
  d^{MNK} \partial_K{\cal O}_{\mu\nu N} \ = \ 0\;,
  \label{conO}
 \ee 
that drop out from (\ref{defF}).
 The  field strengths of vector and 2-form tensor fields are related by
the  Bianchi identities
 \bea
  3\, {\cal D}_{[\mu}{\cal F}_{\nu\rho]}{}^M &=& 10\, d^{MNK}\partial_K{\cal H}_{\mu\nu\rho\,N}\;,
  \nonumber\\
4\, {\cal D}_{[\mu} {\cal H}_{\nu\rho\sigma]M} &=& 
-3 \,d_{MKL} {\cal F}_{[\mu\nu}{}^K {\cal F}_{\rho\sigma]}{}^L
+\dots\;,
\eea
up to terms annihilated by the projection with $d^{MNK}\partial_K$.

We are now in a position to introduce all the fields of the theory (\ref{finalaction})
together with their transformation behavior, and define the various couplings.
Apart from the vector and tensor gauge fields $A_\mu{}^M$, $B_{\mu\nu\,M}$ introduced above,
the theory features scalar fields parametrizing a symmetric group-valued E$_{6(6)}$ matrix ${\cal M}_{MN}$,
transforming as a tensor according to (\ref{deltaV}) w.r.t.\ both indices, and weight $\lambda=0$. Finally,
the f\"unfbein $e_{\mu}{}^{a}$ transforms as a scalar of weight $\lambda=\frac13$ w.r.t.\ (\ref{deltaV}),
i.e.\ carries covariant derivatives according to
\bea
{\cal D}_\mu e_\nu{}^a &\equiv& \partial_\mu e_\nu{}^a - A_\mu{}^M\partial_M e_\nu{}^a
-\frac13\, \partial_MA_\mu{}^M e_\nu{}^a
\,.
\label{covderE}
\eea
The full E$_{6(6)}$-covariant 
field content is thus given by
\bea
\left\{e_\mu{}^a, {\cal M}_{MN}, A_\mu{}^M, B_{\mu\nu\,M} \right\}
\;.
\label{fieldcontent}
\eea

The kinetic terms for scalar and vector fields in (\ref{finalaction}) 
take the manifestly covariant form defined via (\ref{DV}), (\ref{defF}),
with the proper total weight of the integrands adding up to $\lambda=1$, 
as required for invariance of the action.
Next, the topological term $\cal{L}_{\rm top}$ in (\ref{finalaction}) 
is required in order to reproduce the correct duality relations 
between the 1- and 2-forms.
Formally, this topological term is most conveniently defined by considering 
the five-dimensional space as the boundary of a six-dimensional bulk ${\cal M}_6$, 
in which it takes the covariant form of 
a total derivative:
\bea
S_{\rm top} &=& \kappa \!\int d^{27}Y \int_{{\cal M}_6}\,\left(
d_{MNK}\,{\cal F}^M \wedge  {\cal F}^N \wedge  {\cal F}^K\right.
\qquad
\nonumber\\
&&{}\qquad\qquad\qquad
\left.-40\, d^{MNK}{\cal H}_M\,  \wedge \partial_N{\cal H}_K
\right)
\,,
\label{CSlike}
\eea
with a constant $\kappa$ to be determined in the following. 
Alternatively, we may give a non-manifestly covariant explicit expression
for $\cal{L}_{\rm top}$ directly in five dimensions, cf.~\cite{deWit:2004nw}, 
whose variation is given by
\bea
\delta{\cal L}_{\rm top} &=&
\kappa\,\varepsilon^{\mu\nu\rho\sigma\tau}
\Big(\,\frac34\,d_{MNK}\,
{\cal F}_{\mu\nu}{}^M {\cal F}_{\rho\sigma}{}^N  \delta A_\tau{}^K
\nonumber\\
&&{}\qquad
+5\,d^{MNK} d_{KPQ}\,\partial_N{\cal H}_{\mu\nu\rho \,M} \,
A_{\sigma}{}^P\delta A_{\tau}{}^Q
\nonumber\\
&&{}\qquad
+5\,d^{MNK}\,\partial_N{\cal H}_{\mu\nu\rho \,M} \,
\delta B_{\sigma\tau\,K} \Big)
\;,
\label{vartopo}
\eea
from which gauge invariance can explicitly be
checked by means of (\ref{deltaAB}).
The field equations derived by variation of 
$B_{\mu\nu M}$ in (\ref{finalaction}) imply the duality relations 
 \be\label{dualityrel}
d^{PML}\partial_L  \left(e{\cal M}_{MN} {\cal F}^{\mu\nu N}
 +\kappa  \varepsilon^{\mu\nu\rho\sigma\tau}\,
  {\cal H}_{\rho\sigma\tau M}\right) = 0
  \,,
  \ee
showing that, as required, the 2-forms do not represent independent physical degrees of freedom.

The first term in (\ref{finalaction}) is the (covariantized) Einstein-Hilbert 
term for the f\"unfbein $e_{\mu}{}^{a}$. 
It results from the conventional Einstein-Hilbert term by 
introducing covariant derivatives according to (\ref{covderE}) 
and defining the improved Riemann tensor, 
 \be
  \widehat{R}_{\mu\nu}{}^{ab}  =  R_{\mu\nu}{}^{ab}+{\cal F}_{\mu\nu}{}^{M}
  e^{a}{}^{\rho}\partial_M e_{\rho}{}^{b}\,,
 \ee
which is necessary for local Lorentz invariance in presence of (non-commuting) 
covariant derivatives ${\cal D}_{\mu}$~\cite{Hohm:2013nja}.  The final 
term in (\ref{finalaction}) is the `scalar potential'  given by 
\bea\label{pot}
  V &= &\frac{1}{24}\, {\cal M}^{MN}\partial_M{\cal M}^{KL}
  \left(12\,\partial_L{\cal M}_{NK}
  -\partial_N{\cal M}_{KL}\right)\nonumber\\
  &&
  -e^{-1}\partial_Me\,\partial_N{\cal M}^{MN}- {\cal M}^{MN}e^{-1}\partial_Me\,e^{-1}\partial_Ne\nonumber\\
&&  -\frac{1}{4}{\cal M}^{MN}\partial_Mg^{\mu\nu}\partial_N g_{\mu\nu}
\;.
 \eea
Its invariance under generalized diffeomorphisms (\ref{deltaV}) is not manifest
but can be confirmed by an explicit computation.
In particular, we stress that despite the seemingly wrong weight of $\frac53$ carried
by the f\"unfbein determinant $e$ multiplying the potential, the density terms from
the remaining non-covariant contributions of its variation contribute such as to 
guarantee 
$\delta_\Lambda \big(eV\big) = \partial_M \big( e \Lambda^M V\big)$
and thus invariance of the action.
We note that in the truncations of \cite{Berman:2011jh}
the potential is the only term left. While 
the terms ${\cal O}({\cal M}^3)$ in (\ref{pot}) 
can be matched with those in \cite{Berman:2011jh}, comparison 
is subtle beyond that  \footnote{More precisely, the additional terms in \cite{Berman:2011jh} all vanish 
for $\det{\cal M}=1$, i.e., if ${\cal M}$ is a proper group element.}.

The action (\ref{finalaction}) is invariant under all local symmetries discussed above,
which in turn fixes the form of its five separate terms.
Explicit
calculation shows that the action possesses an additional
gauge invariance with parameters $\xi^{\mu}(x,Y)$,
generalizing the standard 5-dimensional diffeomorphisms.
Since these parameters may depend on $Y^M$, this symmetry is non-manifest. Rather, it 
uniquely determines the relative coefficients in (\ref{finalaction})
and in particular $\kappa^2 = \frac5{18}$
for the constant from (\ref{vartopo}). 
This symmetry acts on the fields as 
\bea
\delta e_{\mu}{}^{a} &=& \xi^{\nu}{\cal D}_{\nu}e_{\mu}{}^{a}
+ {\cal D}_{\mu}\xi^{\nu} e_{\nu}{}^{a}\,, \nonumber\\[0.5ex]
\delta {\cal M}_{MN} &=& \xi^\mu \,{\cal D}_\mu {\cal M}_{MN}\,,\nonumber\\[0.5ex]
\delta A_{\mu}{}^M &=& \xi^\nu\,{\cal F}_{\nu\mu}{}^M + {\cal M}^{MN}g_{\mu\nu}  \partial_N \xi^\nu
\;,\nonumber\\[0.5ex]
\delta B_{\mu\nu M} &=& \tfrac1{16\kappa}\,\xi^\rho 
 e\varepsilon_{\mu\nu\rho\sigma\tau} {\cal M}_{MN}  {\cal F}^{\sigma\tau N} \nonumber\\
   &&{}-d_{MKL}\,A_{[\mu}{}^K\, \delta A_{\nu]}{}^L
 \,,
 \label{skewD}
\eea
and for the first three fields resembles standard diffeomorphism transformations, 
albeit with covariant derivatives and an extra contribution in $\delta A_{\mu}{}^M$ \cite{Hohm:2013jma,Hohm:2013nja}.
Remarkably, for the two-forms $B_{\mu\nu M}$, off-shell invariance of the action 
requires that their standard transformation under (covariant) diffeomorphisms is modified by 
a contribution proportional to the duality equations (\ref{dualityrel}).

We have thus constructed with (\ref{finalaction}) 
the unique E$_{6(6)}$-covariant two-derivative action with field content (\ref{fieldcontent}) and symmetries
(\ref{deltaV}), (\ref{deltaAB}), (\ref{skewD}). Moreover, we note that upon genuine dimensional reduction,
i.e.\ setting $\partial_M = 0$, this theory reduces to the E$_{6(6)}$ invariant form
of the maximal five-dimensional theory. In contrast, in (\ref{finalaction}) all fields in addition
depend on the 27 coordinates $Y^M$ modulo the condition (\ref{sectioncondition}).
The latter can be locally solved upon breaking E$_{6(6)}$ under SL$(6)\,\times\, $SL$(2)$ such that
\be
{\bf 27} ~\rightarrow~ (15,1) + (6,2)
\;,
\label{27}
\ee
and having fields depend on only 6 coordinates from the SL$(2)$ doublet.
We will now show how with this choice the action  
(\ref{finalaction}) reproduces the fields and dynamics of 
the conventional form of the full $D=11$ supergravity. 
The bosonic fields of $D=11$ supergravity are the elfbein $E_{\hat{\mu}}{}^{\hat{a}}$ and 
the 3-form potential $C_{\hat{\mu}\hat{\nu}\hat{\rho}}$, where $\hat{\mu}, \hat{\nu}=0,\ldots,10$, 
and $\hat{a}, \hat{b}=0,\ldots, 10$, denote $D=11$ curved and flat indices, respectively. 
Their action reads 
 \bea
   S_{11} &=& \int d^{11}x\,E\Big(R-\frac{1}{12}F^{\hat{\mu}\hat{\nu}\hat{\rho}\hat{\sigma}}F_{\hat{\mu}\hat{\nu}\hat{\rho}\hat{\sigma}}
  \label{D11}\\[0.5ex]
  &&{}\hspace{-0.1cm}
  +\frac{1}{12\cdot 216}E^{-1}\varepsilon^{\hat{\mu}_1\cdots \hat{\mu}_{11}}F_{\hat{\mu}_1\cdots \hat{\mu}_4}
  F_{\hat{\mu}_5\cdots \hat{\mu}_8}C_{\hat{\mu}_9\hat{\mu}_{10}\hat{\mu}_{11}}\Big)\,, 
\nonumber
 \eea
with the abelian field strength
$F_{\hat{\mu}\hat{\nu}\hat{\rho}\hat{\sigma}} = 4\partial_{[\hat{\mu}}C_{\hat{\nu}\hat{\rho}\hat{\sigma}]}\,.$  
For our purpose it turns out to be convenient to work with the on-shell equivalent form of
$D=11$ supergravity in which we add the dual 6-form potential 
$C_{\hat{\mu}_1\cdots \hat{\mu}_6}$ and supplement the field equations 
following from (\ref{D11}) by the duality relation 
 \be\label{dualityrel11}
  F_{\hat{\mu}_1\cdots \hat{\mu}_7} =  \frac{1}{4!}\,E\,\varepsilon_{\hat{\mu}_1\cdots \hat{\mu}_7
  \hat{\nu}_1\ldots\hat{\nu}_4} F^{\hat{\nu}_1\ldots\hat{\nu}_4}\,,
 \ee 
where $F_{\hat{\mu}_1\cdots \hat{\mu}_7} \equiv 7\partial_{[\hat{\mu}_1}C_{\hat{\mu}_2\cdots\hat{\mu}_7]}
   +35 C_{[\hat{\mu}_1\cdots\hat{\mu}_3} F_{\hat{\mu}_4\cdots \hat{\mu}_7]}$\,.
We now perform a $5+6$ splitting of the coordinates, $x^{\hat{\mu}}=(x^{\mu},y^{m})$, and 
partially gauge fix the local Lorentz group $SO(1,10)$ to $SO(1,4)\times SO(6)$, splitting the 
flat indices as $\hat{a}=(a,\alpha)$, by 
taking the elfbein to be of the standard Kaluza-Klein form 
 \be
  E_{\hat{\mu}}{}^{\hat{a}} = \left(\begin{array}{cc} \phi^{-\frac{1}{3}}e_{\mu}{}^{a} &
  A_{\mu}{}^{m} \phi_{m}{}^{\alpha} \\ 0 & \phi_{m}{}^{\alpha}
  \end{array}\right)\,, 
 \ee 
with $\phi=\det (\phi_{m}{}^{\alpha})$. Again, we stress that 
here this ansatz does not entail any truncation 
in that the fields still depend on all $5+6$ coordinates.  
Next, we perform redefinitions of all form fields 
originating from $C_{(3)}$ and $C_{(6)}$ in the usual 
Kaluza-Klein manner, in particular flattening the world indices with the elfbein 
$E_{\hat{a}}{}^{\hat{\mu}}$ and then `un-flattening' with the f\"unfbein 
$E_{\mu}{}^{a}$. The resulting fields are denoted by $A$. 
In total the scalars are given by 
 \be
  {\cal M}_{MN}\,: \, (\phi_{mn}\,,\; A_{mnk}\,,\; A_{mnklpq})\,:\;\quad 21+20+1\,, 
  \label{D5M}
 \ee
giving rise to the $42$ scalars encoded in the group element ${\cal M}_{MN}$ parametrizing 
the coset space E$_{6(6)}/$USp$(8)$, as made explicit in~\cite{Cremmer:1997ct},
however still depending on all eleven coordinates.
Similarly, the E$_{6(6)}$ vector field components originate from the Kaluza-Klein 
vector and from the 3- and 6-form:
 \be
  A_{\mu}{}^{M}\,: \,  (A_{\mu}{}^{m}\,,\; A_{\mu\, mn}\,,\; A_{\mu\, mnklp})\,:\;\quad 6+15+6\,,
  \label{D5A}
 \ee
giving precisely the gauge vectors in the $\overline{\bf 27}$ of  E$_{6(6)}$. 
Again, the counting is well known from dimensional reduction of the theory,
but the full structure here is such that upon keeping the full eleven-dimensional
coordinate dependence, the theory takes the exact form of (\ref{finalaction})  
with the gauge fields (\ref{D5A}) now attached to the non-abelian structure
of the generalized diffeomorphisms (\ref{deltaV}).
Finally, the 2-forms originating from the 3- and 6-form are 
 \be
  (A_{\mu\nu\, m}\,,\; A_{\mu\nu\, mnkl})\,: \quad 6+15\,,
 \label{D5B}
 \ee
which, via (\ref{dualityrel11}),  are on-shell dual 
to the corresponding vector fields in (\ref{D5A}). 
Naively, we seem to be missing six 2-forms to account for the 2-forms
in the ${\bf 27}$ of E$_{6(6)}$ entering the covariant formulation above. However, this is precisely consistent with the 
above observation of an additional shift gauge freedom ${\cal O}_{\mu\nu\,M}$ in
(\ref{deltaAB}) which due to (\ref{conO}) and after explicit solution of the section
condition allows to eliminate 6 of the 27 2-forms. Let us note that these 2-forms
missing in (\ref{D5B}) are precisely those dual to the Kaluza-Klein vector fields 
in (\ref{D5A}), i.e.\ those that descend from the $D=11$ dual graviton.

Summarizing, with this solution of the section condition (\ref{sectioncondition}), the field content
of (\ref{finalaction}) matches the field content of $D=11$ supergravity,  
on-shell and up to pure gauge modes, in a $5+6$ splitting of fields  and coordinates. 
Explicit calculation shows that (\ref{finalaction}) precisely 
reproduces the field equations implied by (\ref{D11}) and (\ref{dualityrel11}) 
for these fields with the full eleven-dimensional coordinate dependence.
E.g.\ the duality equations descending from (\ref{dualityrel11}) for the components of
$A_\mu{}^M$ and $B_{\mu\nu\,M}$, descend from (\ref{dualityrel}), etc.
The technical details of this proof go parallel to~\cite{Hohm:2013nja,Hohm:2013jma}
and will be presented in more detail elsewhere~\cite{HohmSamtleben}. 

Another (inequivalent) solution of the section condition~(\ref{sectioncondition}) breaks
(\ref{27}) further to SL$(5)$ and has all fields depend only on the 5 coordinates within the 15.
The resulting 10-dimensional theory with unbroken SL$(2)$ can be nothing else 
but the IIB theory; indeed all of the above discussion goes through in perfect analogy.

After having discussed the E$_{6(6)}$ case in some detail let us briefly comment on the 
remaining (finite-dimensional) groups. First, the  E$_{7(7)}$ case corresponding to 
the $4+7$ splitting of $D=11$ supergravity can be straightforwardly obtained along the 
lines discussed here, with the action (\ref{finalaction}) replaced by an action of the form
of non-abelian gauged maximal supergravity \cite{de Wit:2007mt}. 
E.g.\ the field strengths for the 56 vector fields take the form
\bea
{\cal F}^{\circ}_{\mu\nu}{}^M \equiv
 2\partial_{[\mu} A_{\nu]}{}^M 
-\big[A_{\mu}, A_{\nu}\big]^M_{\rm E}
+ \Omega^{MK} (t_\alpha)_{K}{}^{N} \partial_N B_{\mu\nu}^{\alpha}\,, 
\nonumber
\eea
with the corresponding bracket from~\cite{Berman:2012vc}, the symplectic tensor $\Omega^{MN}$, 
and 133 2-forms in the adjoint representation of E$_{7(7)}$.
For the fully E$_{7(7)}$-covariant form of the action, the latter needs to be supplemented
by a self-duality constraint between the gauge vectors due to electric-magnetic duality in the $D=4$ language. 
Moreover the appearance of components from the $D=11$ dual graviton among the 56 vector fields
necessitates the proper implementation of the mechanism of constrained compensator fields,
introduced in \cite{Hohm:2013jma} for the case of pure gravity.
In the E$_{8(8)}$ case, the very same mechanism also circumvents the seeming obstacle of non-closure of 
the algebra of E$_{8(8)}$ generalized diffeomorphisms (\ref{deltaV})~\cite{Berman:2012vc}.
Details will appear in~\cite{HohmSamtleben}.

We have restricted the present construction to the bosonic sector, but we are confident that
the extension to include fermions and explicit supersymmetry, as in \cite{Coimbra:2012af}, is straightforward 
in accordance with the structure of the corresponding gauged supergravities in the embedding tensor
formalism~\cite{deWit:2004nw,de Wit:2007mt}. One may expect to further incorporate
higher-derivative M-theory corrections along the lines of~\cite{Hohm:2013jaa}.
Among the most intriguing questions raised by our construction is 
the possibility of finding solutions of the section conditions (\ref{sectioncondition}) that only locally match standard supergravity but 
that globally require a generalized notion of manifold, as happens in DFT \cite{Hohm:2013bwa}. 
This would genuinely transcend the realm of $D=10$ and $D=11$ supergravity, as would a 
relaxation of the section conditions themselves.

\noindent\textbf{Acknowledgements}:  
We would like to thank each others home institutions for hospitality. 
OH is supported by the DFG Transregional Collaborative Research Centre and the DFG cluster of excellence ``Origin and Structure of the Universe''.

\end{document}